\documentclass[aip,pof,amsmath,amssymb,preprint,pof,
]{revtex4-1}

\usepackage{graphicx}
\usepackage{dcolumn}
\usepackage{bm}

\usepackage[utf8x]{inputenc}
\usepackage{etoolbox}
\usepackage{tikz}
\graphicspath{{plots/}}

\makeatletter
\def\@email#1#2{%
 \endgroup
 \patchcmd{\titleblock@produce}
  {\frontmatter@RRAPformat}
  {\frontmatter@RRAPformat{\produce@RRAP{*#1\href{mailto:#2}{#2}}}\frontmatter@RRAPformat}
  {}{}
}%
\makeatother
\begin{document}

\preprint{AIP/123-QED}

\title[]{Interaction of vortex stretching with wind power fluctuations}
\author{Jahrul Alam}
 \email{alamj@mun.ca}
 \affiliation{ 
  Mathematics and Statistics, Memorial University of Newfoundland
}%


\date{\today}

\begin{abstract}
The transfer of turbulence kinetic energy from large to small scales occurs through vortex stretching. Also, statistical properties of the subgrid-scale energy fluxes depend on the alignment of the vorticity vector with the principal strain axis.  A heuristic analysis of the present study indicates that vortex-stretching and the second invariant of the velocity gradient tensor provide a scale-adaptive parameterization of the subgrid-scale stresses and the local energy fluxes in the wakes of wind turbines. The scale-adaptivity underlies the restricted Euler dynamics of the filtered motion that vortex-stretching plays in the growth of the second invariant of filtered velocity gradient and the local energy transfer. We have analyzed wind power fluctuations in a utility-scale wind farm with $41$ actuator disks. The numerical results show that the spectrum of the wind power fluctuations follows a power law with a logarithmic slope of $−5/3$. Furthermore, POD analysis indicates that the wind power fluctuations depend on the incoming turbulence and its modulation by the wake interactions in wind farms.
\end{abstract}

\maketitle

\section{Introduction}
The interaction between wind farms and the atmospheric boundary layer produces fine-scale vortices~\cite{Calaf2010,Abkar2016,Xie2017,Stevens2017,Forooghi2020}. Turbulence kinetic energy (TKE) is progressively moved toward small scale flow features when a vortex stretches~\cite{Taylor38,Johnson2021,Hossen2021}. Representing the nature of energy cascade in atmospheric computational models is vital for understanding the impact of atmospheric turbulence in large wind farms~\cite{Keith2004,Roy2004}. The overall impact of unresolved fine-scale flow on wind power fluctuations is known, but not fully explored through atmospheric turbulence theory~\cite{Bandi2017,Bossuyt2017}. The thrust of this paper is an exposition of how the vortex stretching mechanism connects to turbulence dissipation in the wakes of large wind farms. 

\subsection{Background}
Some studies~\cite{Politis2012} observed a slower wake recovery for wind turbines on the top of hills; however, other studies~\cite{Hyvarinen2017} observed a faster wake recovery. Existing reports on the seasonal variability for the meteorological impact of large wind farms are inconsistent~\cite{Roy2004,Keith2004,Archer2013b,Archer2013,Archer2018}. The literature disagrees on the deflection of wakes caused by the Coriolis force~\cite{Abkar2016,Laan2017}.  It is thus a challenging endeavor to conduct a parametric case study of the direct influence of vortex stretching on wind farms. Suitable characterizations of such phenomena are of great importance in designing strategies for increased penetration of renewable projects within the global energy mix~\cite{Roy2011}, which can aid in wind farm layout optimization and control, as well as understanding the inherent variability of the aerodynamic power of wind farms~\cite{Calaf2010,Roy2011,Bossuyt2017,Xie2017,Stevens2017,Alam2019}.

\citet{Liu2021} observed that surface protuberance could increase the aerodynamic power of wind farms; however, the optimal way of controlling roughness elements of complex terrain is not fully clear~\cite{Alam2020}. In a wind tunnel experiment with five wind turbines sited over hilly terrain, each with a rotor diameter of $25.4$~cm, \citet{Tian2013} observed that the efficiency of wind turbines improved significantly due to local speed-up effects. \citet{Bossuyt2017} presented the measured velocity field over an array of $100$ micro turbines with a rotor diameter of $3$~cm. The outcome suggests that the power fluctuations follow the $-5/3$ spectrum of inertial range turbulence (see also the work of~\citet{Bandi2017}). Knowing the spectral bound helps estimate the ancillary reserves and influence of large-scale atmospheric turbulence.  

\subsection{Motivation and objective}
This research investigates how the variability of atmospheric boundary layer flow over a complex terrain can influence power fluctuations in wind farms. Extensive experimental studies are being performed to understand the transfer of kinetic energy by atmospheric turbulence and the effects of ground roughness in the design of wind farms~\cite{Tian2013,Bossuyt2017,Chamorro2011}. 
However, challenges include the availability of fewer measurement data and even fewer numerical studies of the mechanism that transfers kinetic energy from the largest-scale motions, where it is produced, to the smallest-scale activities, where it dissipates. 
Most importantly, the dynamical characteristics of energy cascade~\cite{Carbone2020,Hossen2021}, especially for the efficient reduced-order representation of turbulence in practical applications, such as optimal design of wind farms, is of high importance.

Our recent works~\cite{Alam2011,Alam2018,Alam2020} developed immersed boundary methods for large eddy simulation (LES) of orographic influence in the atmospheric boundary layer. Unlike classical LESs using the Smagorinsky model, which considers only the `strain portion' of the velocity gradient tensor, our approach considers both the `strain tensor' and the `rotation tensor' to model the eddy viscosity~\cite{Hossen2021}. This approach allows us to dynamically adapt the energy dissipation rates to the scale of the energetic eddies in the roughness sublayer. Note that the LES technique developed by~\citet{Deardorff72,Deardorff70} at the National Center for Atmospheric Research has become a prominent tool for simulating the effects of surface protrusions on the aerodynamic power of wind turbines~\cite{Liu2021}. For example, suppose computational resources permit many grid points to resolve wind turbines' tip vortices. Then the choice of a more sophisticated ``mixed model" over the standard ``Samgorinsky model" may not be the primary criteria for determining simulation accuracy. \citet{Singh2022} proposed an improved actuator disk model for wind farm simulations, and assessed the method's performance using wind tunnel measurements. The current article presents a scale-adaptive LES of utility-scale wind farms considering an array of $41$ RE-power 5-MW wind turbines~\cite{Alam2019}. It focuses on relating spectral features of wind turbines' wakes to properties of atmospheric boundary layer flows over complex terrain~\cite{Chow2010,Alam2011,Chow2019,Alam2020}. 

This article focuses on two approaches for the numerical simulation of atmospheric turbulence around wind farms. First, we consider the Navier-Stokes (NS) equations linearized around the mean profile of the atmospheric boundary layer flow~\cite{Zare2020}. The main idea is to leverage the underlying physics in the form of a prior model that arises from first principles, such as the NS equations~\cite{Kevlahan2004b}. Stochastically forced linearized dynamics transfers energy between the mean flow and velocity fluctuations. For instance, like in stochastic climate prediction models, white-in-time stochastic forcing of linearized NS equations lead to quasi-geostrophic turbulence~\cite{Hirt2019}. This approach is beneficial for directly observing the interaction between the array of wind turbines and the incoming atmospheric turbulence.

Second, we investigate the scale-adaptivity of dissipation rates~\cite{Kurowski2018} in LES of wind farms, where we utilize the transfer of energy via vortex stretching~\cite{Carbone2020,Johnson2020,Hossen2021}. Note that LES accurately captures the large-scale unsteady motions in wind farms~\cite{Meneveau2017}, using fine grids combined with secondary wall models~\cite{Bose2018}, thanks to the recent progress in computing power. However, sufficiently fine grid resolutions are often not affordable when performing the LES of wind farms with many turbines. In our scale-adaptive LES~\cite{Alam2018,Alam2019,Alam2020}, stretching a vortex is the dynamic mechanism for flow structure evolution and energy cascade, highlighting that regions of enhanced strain surround the elements of enhanced vorticity~\cite{Carbone2020,Johnson2020}. However, a detailed understanding of the mechanisms responsible for energy transfer in turbulent flows remains elusive. Thus, our knowledge of wind farm aerodynamics is primarily empirical, {\em albeit} relying on physical intuition, numerical simulation, and experimental measurements. In the scale-adaptive criteria, filtered strain-rate tensors align perfectly with stresses arising from the low pass filter. 

These observations about turbulence and wind farms suggest a need to explore the potential for scale adaptivity of subgrid-scale turbulence in the LES framework across a full range of resolutions employed by wind farm simulation models. Section~\ref{sec:vst} focuses on developing subgrid-scale transport schemes using the vortex stretching mechanism. Section~\ref{sec:sdm} extends the stochastic dynamical modeling approach as a tool to transfer energy between the mean flow and velocity fluctuations in idealized wind farms. Section~\ref{sec:apf} investigates the spectrum of aerodynamic power fluctuations in a simulated wind farm. Finally, section~\ref{sec:cfd} summarizes the findings of the present research.

\section{Energy cascade and vortex stretching}
\label{sec:vst}
This section studies the technical aspect of a pragmatic scale-adaptive LES model of atmospheric boundary layer flows.  Let $L$ and $\eta$ denote energy-containing and dissipating length scales, respectively. If we apply an isotropic filter at a length-scale $\Delta_{\hbox{\tiny les}}$ such that $L\gg\Delta_{\hbox{\tiny les}}\gg\eta$, the relative alignment of the filtered strain-rate tensor with the subfilter stress tensor estimates the local energy cascade rate across $\Delta_{\hbox{\tiny les}}$, which can close the filtered momentum equations~\cite{Carbone2020}. Moreover, the subfilter stress tensor is a function of the filtered velocity gradient tensor~\cite{Hossen2021}. As a result, the local energy cascade rate depends on the skewness of filtered strain plus the stretching of filtered vorticity~\cite{Eyink2006,Johnson2020}. Energy containing $\mathcal O(L)$ eddies connect the constrained (or restricted) Euler dynamics to the energy cascade~\cite{Trias2015,Hossen2021}. Whereas energy cascading $\mathcal O(\Delta_{\hbox{\tiny les}})$ eddies, characterized by velocity gradients,  extract energy from the large-scale eddies via vortex stretching and pass it onto dissipating eddies of size $\mathcal O(\eta)$~\cite{Eyink2006}. The scale-adaptive LES method thus aims to capture the natural tendency of the hierarchical production of small-scale structures in turbulent flows~\cite{Kolmogorov41,Tsinober2001,Davidson2004}.
%
\subsection{Amplification of enstrophy in turbulent flows}
We now briefly extend exact mathematical results toward a large-scale circulation balance in (atmospheric) turbulence, while considering low-pass filtered velocity fields~\cite{Tsinober2001}. One of the best-known mathematical results of such a circulation balance is due to~\citet{Foias89}. It asserts that the regularity and uniqueness of the velocity $\bm u(\bm x,t)$ are guaranteed up to a finite time if the enstrophy ({\em i.e.} $(1/2)|\bm\nabla\times\bm u|^2$) of the flow remains bounded~\cite{Kang2020}. The mathematical analysis of~\citet{Foias89} is consistent with an earlier experimental result due to~\citet{Taylor38}, which depends upon an exact consequence of the NS dynamics:
\begin{equation}
  \label{eq:enst}
\frac{\partial}{\partial t}\left(\frac12|\bm\omega|^2\right) = \bm\omega^T\bm S\bm\omega - \nu|\bm\nabla\bm\omega|^2.
\end{equation}

Now, consider the filtered NS equations governing the atmospheric boundary layer flow~\cite{Garratt92,Pope2000}. Let $\tau$ represent the cumulative effect of the subfilter-scale stress $\tau^s = [\overline{u_iu}_j - \bar u_i\bar u_j]$ plus the viscous stress $2\nu\bm S$. An evolution equation for filtered enstrophy is (see also Eq.~51 of~\citet{Meneveau94})
\begin{equation}
  \label{eq:ensf}
\frac{\partial}{\partial t}\left(\frac12|\bar{\bm\omega}|^2\right) = \bar{\bm\omega}^T\mathcal S\bar{\bm\omega} - \left[\bm\nabla\cdot(\tau - \frac13\rm{tr}~\tau)\right]\cdot(\bm\nabla\times\bar{\bm\omega})
\end{equation}
where $\mathcal S$ and $\bm S$ represents filtered and exact (or unfiltered) rates of strain, respectively. 
Estimates of finite a priori bounds on the growth of enstrophy $(1/2)|\bm\omega|^2$ (a measure of the vorticity $\bm\omega=\bm\nabla\times\bm u$) remain an open question on the regularity problem for the 3D NS system~\cite{Kang2020}. Wind tunnel measurements suggest that vorticity is produced by vortex stretching three times as fast as it dissipates~\cite{Taylor38}. From Eq~(\ref{eq:ensf}), in the absence of viscous effects in the inertial range, vortex stretching is expected to be highly correlated with turbulence dissipation in order to ensure a finite bound on enstrophy.

\subsection{The dynamic model of subgrid scale turbulence}
Existing dynamic modeling approaches lean indirectly on the fact that the transfer of energy from large to small eddies is a local interaction associated with vortex stretching~\cite{Richardson22,Taylor38,Lumley92,Davidson2004,Doan2018}.  In the physical space, the transfer of energy occurs  through the production of vorticity~\cite{Pope2000,Tsinober2001,Johnson2021,Hossen2021}. The vorticity  of a fluid element may be reoriented or concentrated or diffused depending on the motion of that fluid element and on the torques applied to it by the surrounding fluid elements~\cite{Javier93,Saffman93}. The role of vortex stretching in the turbulence energy cascade has been thoroughly evaluated by~\citet{Johnson2020}; that study also highlights the contribution of strain self-amplification. Indeed, \citet{Batchelor49} ({\em e.g.} pg 253) precisely formulates that an interaction between two neighboring energetic eddies can dissipate energy only on a very small scale~\cite{Tsinober2001}.

Consequently in LES, the subfilter stress tensor $\tau^s_{ij} = \overline{u_iu}_j - \bar u_i\bar u_j$ is typically assumed to be precisely aligned with the eigenframe of the symmetric part of the velocity gradient tensor, $\mathcal S_{ij} = (1/2)(\partial\bar u_i/\partial x_j + \partial\bar u_j/\partial x_i)$. Following~\citet{Smagorinsky}, a model ($\tau_{ij}$) of the subfilter scale stress $\tau^s_{ij}$ is defined as
\begin{equation}
  \label{eq:smg}
  \tau_{ij} -\frac13\delta_{ij}\tau_{kk}= 2\nu_\tau\mathcal S_{ij},
\end{equation}
where the eddy viscosity $\nu_\tau(\bm x,t)$ is prescribed in a way to generate the appropriate flux of the energy, $\tau_{ij}\mathcal S_{ij}$. The dynamic procedure~\cite{Germano91} proposes to compute the Leonard (or resolved) stress 
 \begin{equation}
   \label{eq:tl}
   \tau^L_{ij} = \widetilde{\bar u_i\bar u}_j-\tilde{\bar u}_i\tilde{\bar u}_j.
 \end{equation}
 Eq~(\ref{eq:tl}) filters the resolved velocity $\bar u_i(\bm x,t)$ over a larger box of size $\alpha\Delta_{\hbox{\tiny les}}$ than the kernel of low pass filtering. The eddy viscosity is then computed as
 $$
 \nu_\tau(\bm x,t) = 2c_s(\bm x,t)\Delta^2_{\hbox{\tiny les}}|\mathcal S|.
 $$
 Assuming that the flow is homogeneous ({\em e.g.} channel flow) on a plane over which an average $\langle\cdot\rangle_h$ can be justified, we have
\begin{equation}
  \label{eq:cs}
  c_s(\bm x,t) = \frac{\langle\tau^L_{kl}\mathcal S_{kl}\rangle_h}{\alpha^2\Delta^2_{\hbox{\tiny les}}\langle |\tilde{\mathcal S}|\tilde{\mathcal S}_{mn}\mathcal S_{mj} \rangle_h - \langle|\mathcal S|\mathcal S_{pq}\mathcal S_{pq} \rangle_h}.
\end{equation}
Here, $\tilde{\mathcal S}_{ij} = (1/2)(\partial\tilde{\bar u}_i/\partial x_j + \partial\tilde{\bar u}_j/\partial x_i)$. Modern machine learning algorithms may also gradually improve the estimates of the dynamic parameter $c_s(\bm x,t)$~\cite{Novati2021}. The eddy viscosity may become negative on some grid points due to negative value of $c_s(\bm x,t)$. A technical details of this classical dynamic procedure is given by~\citet{Germano91}. 

If the velocity gradient field were to characterize the small-scale eddies, it is reasonable to assume that the eddy viscosity $\nu_\tau(\bm x,t)$ depends on the velocity gradient tensor {\em via} a highly nonlinear relation $\tau^s_{ij} = \pi_\nu(\nu_\tau,\mathcal G_{ij})$, where the resolved velocity gradient tensor $\mathcal G_{ij} = {\partial\bar u_i}/{\partial x_j}$ is the input, and residual stress tensor $\tau^s_{ij}$ is the output~\cite{Brunton2020,Novati2021}. Before discussing the velocity gradient interactions for a scale-adaptive LES, it may help readers if we refer to some of the recent developments in machine learning approaches to turbulence modeling. \citet{Novati2021} propose a reinforcement learning framework for the automated discovery of $\pi_\nu$ using the available flow field data from experiments, field measurements, and direct numerical simulations~(DNS). For instance, $\tau^{\hbox{\tiny DNS}}_{ij} = \overline{u_iu}_j- u_iu_j$ (computed from DNS field $u_i$) becomes the target in a reinforcement learning framework that predicts $\nu_\tau(\bm x,t)$ from a snapshot of the tensor field $\mathcal G_{ij}$. The goal of such data-oriented methods is to fill in the serious need to calculate turbulent flows for the design of all sorts of devices, where we need relatively cheap and reliably accurate ways for the computation of turbulent flows. 

\subsection{A subgrid model based on vortex stretching}
The scale-similarity hypothesis, due to~\citet{Leonard74}, states that the subfilter-scale stress $\tau^s_{ij}$ equals the resolved stresses at scales between $\Delta_{\hbox{\tiny les}}$ and $\alpha\Delta_{\hbox{\tiny les}}$, where Eq~(\ref{eq:tl}) defines the Leonard (or resolved) stress. Past investigators compared the expression in Eq~(\ref{eq:tl}) with the true stress from DNS data~\cite{Borue98}. The resulting correlation coefficient (of order $90$\% or higher) indicates that the Leonard stress $\tau^L_{ij}$ correlates very well with the true subfilter scale stress $\tau^s_{ij}$; however, the usage of Eq~(\ref{eq:tl}) as a subgrid model leads to energy build-up at small scales~\cite{Moser2021}. 

For the resolved velocity $\bar u_i(\bm x,t)$ (that is filtered over a box of size $\Delta_{\hbox{\tiny les}}$), take an average over a larger box of size $\alpha\Delta_{\hbox{\tiny les}}$. Applying the Taylor expansion~\cite{Borue98,Meneveau2000}, 
$$
\bar u_i(\bm x,t)\approx\tilde{\bar u}_i(\bm r,t) + (\bm x-\bm r)\mathcal G_{ij},
$$
a leading order approximation to the Leonard stress~\cite{Borue98,Eyink2006} is
$
\tau^L_{ij} = c_k\Delta^2_{\hbox{\tiny les}}\mathcal G_{ik}\mathcal G_{jk}.
$
Note that the second invariant of the deviatoric Leonard stress
$$
\tau_{ij}^{L^{dev}} = \frac12\left(\tau^L_{ij}+\tau^L_{ji}\right) - (1/3)\tau^L_{kk}\delta_{ij},
$$
is $\mathcal Q_\mathcal L = -(1/2)\tau^{L^{dev}}_{ij}\tau^{L^{dev}}_{ji}$. After some algebraic manipulation (see~\citet{Davidson2004}), we have the following expression
\begin{equation}
  \label{eq:dlt}
  \mathcal Q_\mathcal L = -\frac14\left[\mathcal S_{ij}\omega_j\mathcal S_{ik}\omega_k+\frac13(\mathcal G_{ij}\mathcal G_{ji})^2\right],
\end{equation}
where the last term in the above expression represents the second invariant $\mathcal Q_\mathcal G=-(1/2)\mathcal G_{ij}\mathcal G_{ji}$ of the velocity gradient tensor.

It is interesting to consider the asymptotic limit of the restricted Euler dynamics~\cite{Borue98}, where $(27/4)\mathcal R^2_\mathcal G + \mathcal Q^2_\mathcal G=0$. Now, compare the third invariant of the velocity gradient tensor,
$$
\mathcal R_{\mathcal G} = -\frac13\left[\mathcal S_{ij}\mathcal S_{ik}\mathcal S_{ki} + \frac34\omega_i\omega_j\mathcal S_{ij}\right]
$$
with the energy flux associated with Leonard stress, Eq~(\ref{eq:tl}),
$$
\Pi:=-\mathcal S_{ij}\tau^L_{ij} = c_k\Delta^2_{\hbox{\tiny les}}\left[-\mathcal S_{ij}\mathcal S_{ik}\mathcal S_{ki} + \frac14\omega_i\omega_j\mathcal S_{ij}\right].
$$
The asymptotic limit of the restricted Euler dynamics may be interpreted as the existence of a functional that depends on $\tau^L_{ij}$ and represents the local rate of dissipation in turbulent flows. A negative skewness of strain, $\mathcal S_{ij}\mathcal S_{ik}\mathcal S_{ki}$, along with a positive value of vorticity stretching, $\omega_i\omega_j\mathcal S_{ij}$, will decrease $\mathcal R_\mathcal G$ and increase $\mathcal Q_\mathcal G$. It is thus evident that the stretching of vorticity would extract energy as large-scale strain is enhanced~\cite{Borue98,Johnson2020,Hossen2021}.  Hence, an appropriate functional of $\tau^L_{ij}$ -- may be in the form of its second invariant -- can account for the turbulence dissipation rate {\em via} the vortex stretching mechanism.


To reach in the aforementioned destination, we follow \citet{Deardorff72}, where a global quantity $k_{\hbox{\tiny sgs}}(\bm x,t)$ was proposed for the following form of the subgrid model 
\begin{equation}
  \label{eq:sam}
  \tau_{ij} - \frac13\tau_{kk}\delta_{ij} = c_k\Delta_{\hbox{\tiny les}}\sqrt{k_{\hbox{\tiny sgs}}}\mathcal S_{ij}.
\end{equation}
%
%
This model is widely used by the atmospheric science community, where a transport equation is solved for $k_{\hbox{\tiny sgs}}$. The idea is to ensure that the eddy viscosity $\nu_\tau(\bm x,t)=c_k\Delta_{\hbox{\tiny les}}\sqrt{k_{\hbox{\tiny sgs}}(\bm x,t)}$ is dynamically adjusted while the parameter $c_k$ is fixed.

Based on dimensional reasoning, we form a functional that maps the space of velocity gradient tensor to the field of turbulence kinetic energy, which reads in the following expression of $k_{\hbox{\tiny sgs}}$ in the context of Eq~(\ref{eq:sam})~(see~\citet{Alam2020,Hossen2021}):
\begin{equation}
  \label{eq:ksgs}
  k_{\hbox{\tiny sgs}} = \frac{\Delta^2_{\hbox{\tiny les}}(\frac12\mathcal S_{ij}\omega_j\mathcal S_{ij}\omega_k + \frac16(\mathcal G_{ij}\mathcal G_{ij})^2)^3}{\left[(\mathcal S_{ij}\mathcal S_{ij})^{5/2}+( \frac12\mathcal S_{ij}\omega_j\mathcal S_{ij}\omega_k + \frac16(\mathcal G_{ij}\mathcal G_{ij})^2 )^{5/4}\right]^2}.
\end{equation}
The arrival at Eq~(\ref{eq:ksgs}) is a `heuristic' derivation, numerically examined by our recent work~\cite{Alam2015,Alam2016,Alam2020,Alam2021}. Several similar investigations, such as those dealing with the statistics of velocity gradient tensor ({\em e.g.,}~\citet{Dallas2013},~\citet{Buxton2017},~\citet{Danish2018}), vorticity dynamics, and vortex stretching in turbulent flows ({\em e.g.}~\citet{Meneveau94,Borue98,Nicoud99,Trias2015,Carbone2020,Johnson2020,Johnson2021}) support the mathematical and the numerical analyses involved in the present research.

\subsection{Numerical verification}
\begin{figure}[h]
  \begin{tabular}{c}
    $(a)$ \\
    \includegraphics[width=8cm]{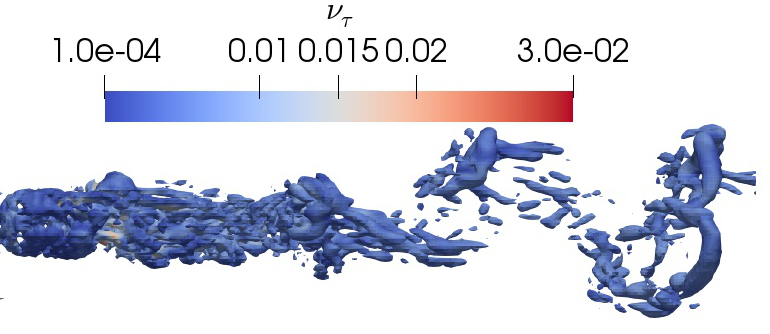}\\
    $(b)$\\
    \includegraphics[width=8cm]{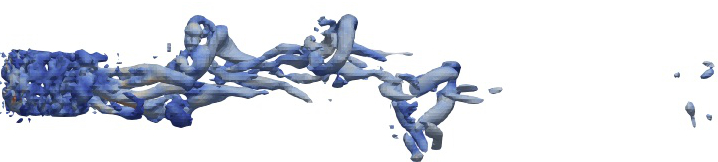}\\
  \end{tabular}
  \caption{\label{fig:spr} Second invariant $Q = (1/2)\mathcal G_{ij}\mathcal G_{ij}$ of the velocity gradient tensor in the wake behind a sphere of fixed diameter, $D=1$~m (for an iso-surface value of $Q=50~\hbox{s}^{-2}$). The color represents a range of the eddy viscosity $\nu_\tau(\bm x,t)$. $(a)$ $\mathcal Re = 3.7\times 10^3$ and $\nu = 2.7\times 10^{-3}$~m$^2$/s; $(b)$ $\mathcal Re = 10^5$ and $\nu = 10^{-4}$~m$^2$/s.}
\end{figure}
\begin{figure}[h]
  \begin{tabular}{c}
    \includegraphics[width=8cm]{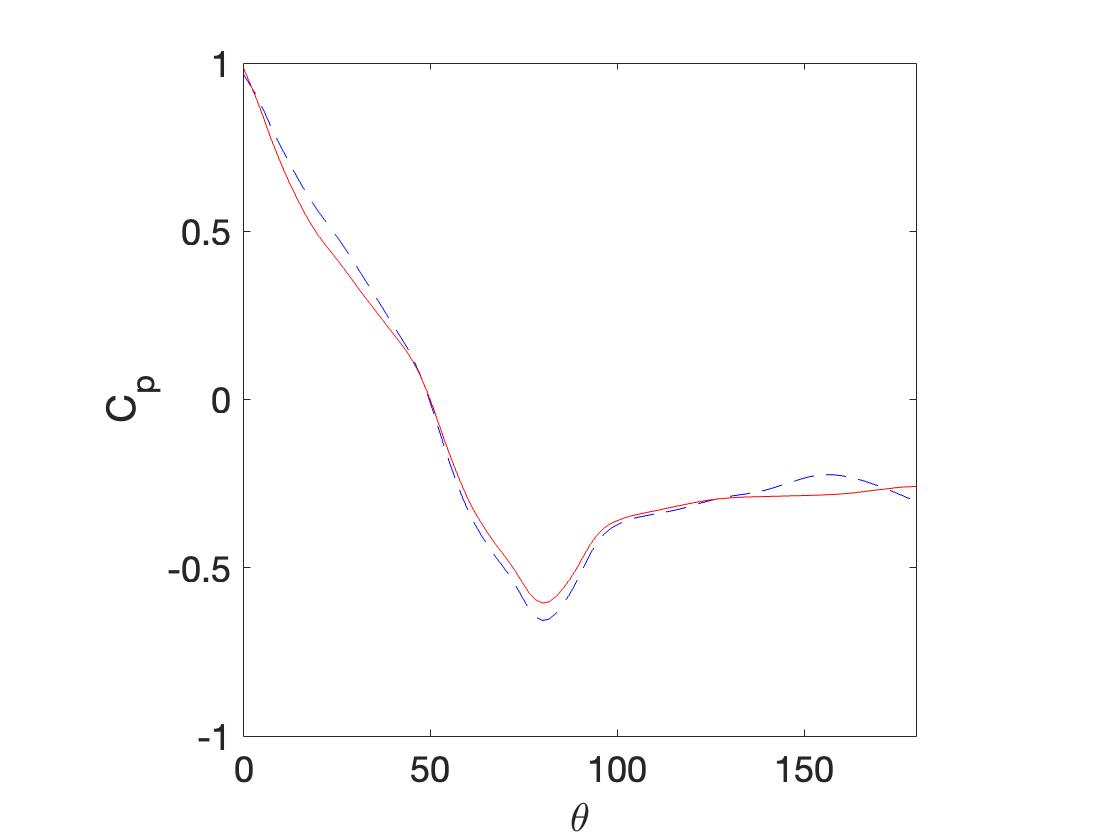}\\
  \end{tabular}
  \caption{\label{fig:cp} The distribution of normalized mean pressure for $\mathcal Re=3.7\times 10^3$ (broken line) and $\mathcal Re=10^5$ (solid line).}
\end{figure}
In order to test the scale-adaptive LES of the vortex formation, this section briefly investigates the coherent structures in the wake of a sphere. The direct forcing immersed boundary method was considered to represent the sphere. Fig~\ref{fig:spr} shows Q-isosurfaces at two Reynolds numbers: $\mathcal Re = 3.7\times 10^3$ and $\mathcal Re= 10^5$. Here, $\mathcal Re=UD/\nu$. In both the cases, the diameter $D=1$~m  and the reference velocity $U=10$~m/s were fixed. The grid was uniformly refined using only $16$ grid points ({\em i.e.} $\Delta/D = 0.0625$) for every length of $D$. In our LES, such a coarse grid is insufficient to resolve the smallest flow scales, the surface of the immersed solid, and the viscous boundary layer. At $\mathcal Re=3.7\times10^3$, the Kolmogorov length scale in the vicinity of a sphere is approximately $\eta/D = 0.0134$, which would decrease by up to a factor of $12$ at $\mathcal Re = 10^5$. 

In the sub-critical range of $\mathcal Re$, the dynamics of flow behind a sphere does not change significantly; {\em i.e.} the shear layer is laminar at separation from the sphere and the wake is turbulent. Moreover, the drag is due mainly to the pressure distribution on the surface of the sphere, with only a minor contribution from viscous shear stress. Fig~\ref{fig:cp} shows that the mean pressure distribution remains about the same for the two Reynolds numbers considered here. Considering that the grid is insufficient to resolve the viscous boundary layer, the pressure distribution in Fig~\ref{fig:cp} exhibits and excellent agreement with the corresponding outcome from experiments or direct numerical simulation. 

\section{Stochastic dynamical modeling}
\label{sec:sdm}
In this section, we simulate a neutrally stratified atmospheric boundary layer  in which a linear stochastically forced input-output system is employed to inject energy into the system~\cite{Zare2020}. 
We employ the similarity theory for modeling the effects of unresolved viscous sublayer in atmospheric boundary layer. 

\subsection{Stochastic forcing in linearized Navier-Stokes equation}
%

Consider a statistically stationary state of turbulence in which the dynamics of velocity and pressure fluctuations $(\bm u',p')$ around the mean $(\bar{\bm U},\bar p)$ satisfies the linearized NS equations. In the absence of Coriolis effects, the mean  $\bar{\bm U} = [U(z),0,0]^T$ exhibits a logarithmic dependence with distance to the Earth's surface. The linearized model is given by the following stochastically forced linear time-invariant system:
\begin{eqnarray}
  \label{eq:lti}
  \frac{\partial u'}{\partial t} + U(z)\frac{\partial u'}{\partial x} &+ v'\frac{dU}{dz} &= -\frac{\partial p'}{\partial x} + \frac{1}{\mathcal Re}\nabla^2u' + r_u,\nonumber\\
  \frac{\partial v'}{\partial t} + U(z)\frac{\partial v'}{\partial x}&  &= -\frac{\partial p'}{\partial x} + \frac{1}{\mathcal Re}\nabla^2v' + r_v,\\
  \frac{\partial w'}{\partial t} + U(z)\frac{\partial w'}{\partial x}&  &= -\frac{\partial p'}{\partial x} + \frac{1}{\mathcal Re}\nabla^2w' + r_w,\nonumber\\
  0& &= \frac{\partial u'}{\partial x} + \frac{\partial v'}{\partial x} + \frac{\partial w'}{\partial x}.\nonumber
\end{eqnarray}
Here,  $\bm r_u = [r_u,r_v,r_w]^T$ is a stochastic forcing term. Fourier transforms in $x$ and $y$ can parameterize the above system of partial differential equations (in $z$ and $t$) by using the horizontal wavenumbers $\bm k:=(k_x,k_y)$. It is possible to eliminate the pressure by expressing the system in which the state is given by the wall-normal velocity $w$ and the streamwise vorticity ($\omega'_y = \partial_yu'-\partial_xv'$). Thus, the velocity fluctuations at any $\bm k$ is determined by the state vector $[w',\omega'_y]$. Under the assumption of scale separation between the large-scale velocity field and the fluctuating velocity $\bm u'$, the random advection of the streamwise velocity fluctuation $u'$ may be considered  approximately constant in space and time. As a result, the linearized dynamics~(\ref{eq:lti}) becomes a finite-dimensional state-space representation in which the velocity fluctuations at any $x$ is given by the state vector $\Psi = [u',\omega'_x]$ such that
\begin{eqnarray}
  \label{eq:psi}
  \frac{\partial\Psi}{\partial t} &=& \mathcal A\Psi + \mathcal B\bm r_u,\\
  \bm u' &=& \mathcal C\Psi.\nonumber
\end{eqnarray}
Writing the linearized system~(\ref{eq:lti}) in the form of Eq~(\ref{eq:psi}) has the benefit of drawing knowledge from the area of control theory for dynamical systems when numerical models that are based on NS equations aim to capture the dynamics and statistical features of turbulent flows.


Using the Green's identity, we get a divergence-free velocity field induced by a concentrated distribution of vortices with arbitrary orientation by solving the Poisson equation
$$
\nabla^2\bm u' = -\bm\nabla\times\bm\omega'.
$$
The linearized dynamics~(\ref{eq:lti}) with stochastic forcing provide a velocity field induced by an aligned vorticity field such that
$$
\bm u'(\bm x,t) = -\frac{1}{4\pi}\int_{V}\frac{1}{|\bm x-\bm y|}\left[\bm\nabla'\times\bm\omega(\bm y,t)\right]d^3\bm y.
$$

One may choose the volume $V$ to be aligned so that one of its boundary surfaces becomes normal to both the vorticity vector $\bm\omega'(\bm x,t)$ and the direction of background mean flow. The fluctuating velocity is then added to the mean background velocity $\bar{\bm U} + \bm u' = [U(z), v'(y,z), w'(y,z)]$ in a specific region, such as the inflow boundary $x=0$.

In contrast to conventional method of generating synthetic turbulence, (i) our algorithm does not require the solution of full NS equation to re-circulate nonlinear eddy dynamics on embedded inflow domains, (ii) provides analytically known representation of unresolved small-scale instabilities, and (iii) useful for efficient representation of a much wider range of unresolved scales.

\subsection{Atmospheric boundary layer similarity theory}
For very light wind with an equivalent roughness length $z_0 = 0.11\nu/u_*$, the eddy viscosity varies like $\mathcal O(z^3)$ for $z < 50\nu/u_*$.  In conditions with high wind above flexible surface protrusions, such as crops or forests, observations suggest that the roughness length $z_0$ can be dependent on the near surface wind speed, such as
\begin{equation}
  \label{eq:cz0}
  z_0 = \alpha_cu_*^2/g,
\end{equation}
where $\alpha_c$ is referred to as Charnock's constant~\cite{Garratt92} and $u_*$ is the friction velocity. In the roughness sublayer $z \le z_*$, the wind profile also deviates from the logarithmic wind profile given by the Monin-Obukhov similarity theory~\cite{Garratt92,Chow2019}. Typical values of $z_*/z_0$ vary between $10$ and about $150$, so that taking $z_*/z_0=100$ implies that for an open forest ($z_0=0.5$~m), the depth of roughness sublayer is $z_*\approx 50$~m.

The drag coefficient $C_d$ in the roughness sublayer for a neutrally stratified atmospheric boundary layer is
\begin{equation}
  \label{eq:cd}
  C_d = [\kappa/\ln(z_*/z_0-\Psi_M(z_*/z_0)]^2
\end{equation}
where $\Psi_M(z/z_0)$ is an empirical function such that $\Psi_M=0$ for $z/z_* \ge 1$. 
In presence of flexible surface protrusions, the surface boundary condition~\cite{Sullivan94}  
$$
\sqrt{\langle\tau_{13}\rangle^2 + \langle\tau_{23}\rangle^2} + \sqrt{\langle u'w'\rangle^2 + \langle v'w'\rangle^2}= u^2_*
$$
is applied. It assumes that atmospheric boundary layer flows take on the character of a shear-driven channel flow in close proximity of a bounding surface. More specifically in the mean energy equation, the exact viscous dissipation $-2\nu\bm{\mathcal S}_{ij}\bm{\mathcal S}_{ij}$ (note the bold face) must equal the filter-scale energy flux plus the resolved viscous dissipation, $-2\nu\mathcal S_{ij}\mathcal S_{ij}$; {\em i.e.,}
$$
\tau_{ij}\mathcal S_{ij} -2\nu\mathcal S_{ij}\mathcal S_{ij} = -2\nu\bm{\mathcal S}_{ij}\bm{\mathcal S}_{ij}.
$$
Thus, a correction of the eddy viscosity in the vicinity of the surface would account for the correct energy flux and a balance of the production and the dissipation.

Following \citet{Schumann75}, the eddy viscosity is corrected by considering the horizontally-averaged subgrid-scale stresses. Thus, in close proximity of Earth's surface~\cite{Sullivan94}, horizontally-averaged stresses are
$$
\langle\tau_{i3}\rangle \approx 2\left[\langle\gamma(z)\nu_\tau\rangle + \nu_{\hbox{\tiny T}}\right]\frac{\partial\langle u_i\rangle}{\partial z},\quad i=1,\,2,
$$
where the damping factor $\gamma(z)$ switches off to Reynolds-averaged Navier-Stokes (RANS) model with a constant eddy viscosity $\nu_{\hbox{\tiny T}}$. However, as it is pointed out by~\citet{Tennekes76}, the small-scale motions in shear layers will roughly be independent of the mean strain rate, in particular, $\partial\langle u_i\rangle/\partial z$. To deal with small-scale motions through vortex stretching, this work has relaxed the horizontal averaging and calculate $\nu_{\hbox{\tiny T}}$ in correcting the eddy viscosity. Thus, instead of the above formulation, we consider
$$
\left[\tau_{13}^2 + \tau_{23}^2\right]^{1/4}  = C^{1/2}_d u(x,y,z_1)
$$
and find the RANS contribution $\nu_{\hbox{\tiny T}}$ from
$$
\tau_{i3} = 2\left[\nu_\tau(x,y,z_1) + \nu_{\hbox{\tiny T}}\right]\frac{\partial u_i}{\partial z},\quad i=1,\,2.
$$
Here, $z_1$ is the vertical level of grid points adjacent to the bounding surface. Recent work~\cite{Alam2018a,Alam2020,Bhuiyan2020} studied such a profile of near-surface turbulence in atmospheric boundary layer flow over hilly terrain. They observed an excellent agreement between the results of the scale-adaptive LES and the corresponding measurements. Below, we study the $k^{-5/3}$ scaling result in the context of the large scale influence of atmospheric turbulence. 


%
\subsection{Numerical simulation}
Consider a neutrally stratified atmospheric boundary layer flow in the domain $7056~\rm{m}\times3024~\rm{m}\times 756~\rm{m}$, which is discretized with $896\times 384\times 96$ grid points ({\em i.e.}~about $33$ millions), where the isotropic grid spacing is $\Delta x=\Delta y=\Delta z=7.875$~m~($\equiv\Delta$). Using a length scale of $D=126$~m, the computational domain is $[-8D,48D]\times[0,24D]\times[0,6D]$. The flow was initialized with an undisturbed mean profile $\bar{\bm U} = [(u_*/\kappa)\log(z/z_0),0,0]^T$.
\begin{figure}[h]
  \centering
  \begin{tabular}{c}
    \includegraphics[height=3cm]{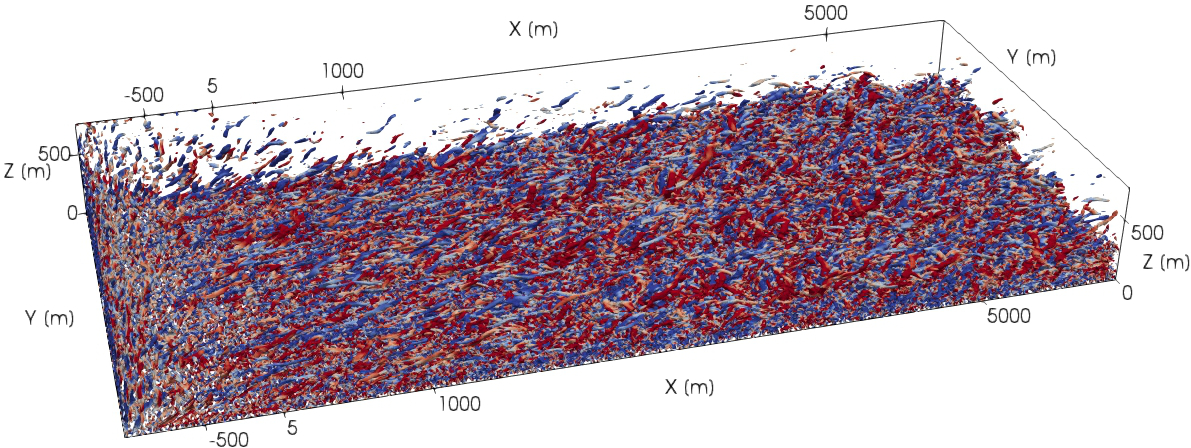}\\
  \end{tabular}
  \caption{\label{fig:q}  A snapshot of the second invariant of the velocity gradient tensor at $t=1500$~s, which is about $130$ units of turnover time for eddies of size $\mathcal O(100$~m). The red and the blue colors represent counter-clockwise and clockwise rotation about $z$-axis. The transition of the inflow condition is seen for $x\le-500$~m.}
\end{figure}
\begin{figure}[h]
  \centering
  \begin{tabular}{c}
    \includegraphics[height=7cm]{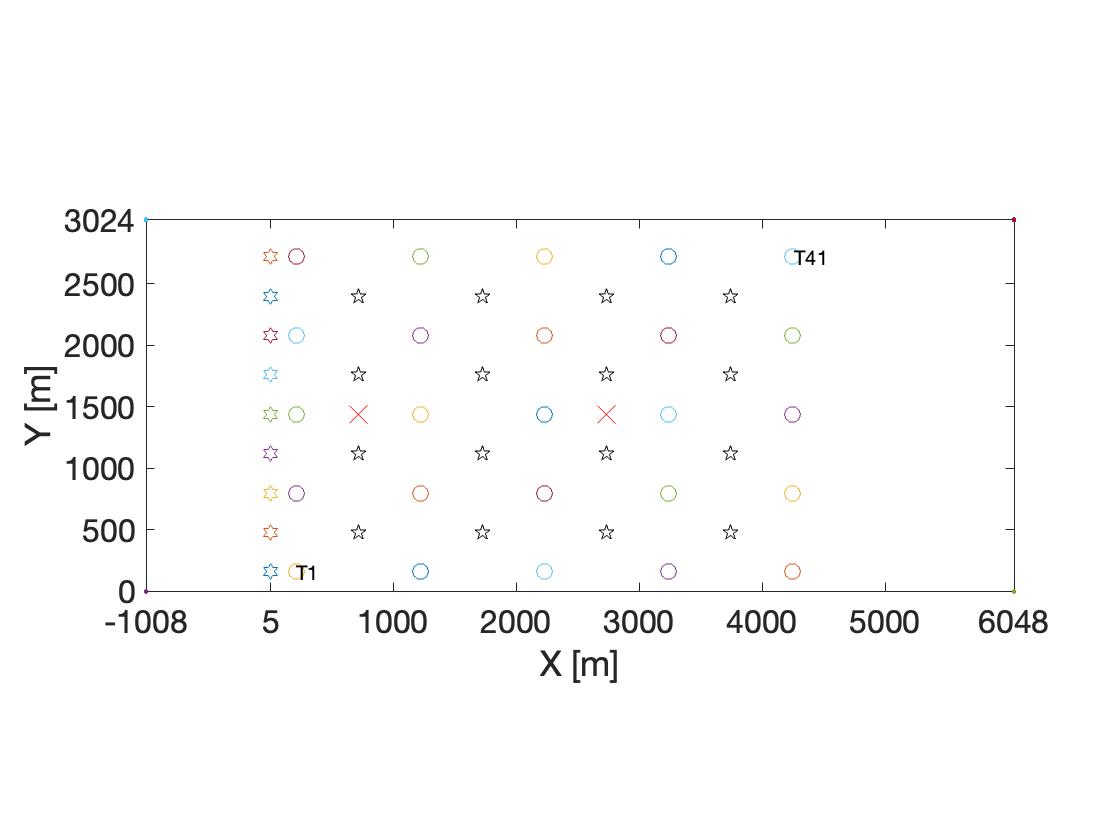}
  \end{tabular}
  \caption{\label{fig:prb} A sketch of the horizontal extent of the computational domain. The velocity signals were collected at $50$ locations. The locations on the front row at $x=5$ are denoted by F1-F9. All other locations (except the two $\times$'s) denoted by T1-T41 denote the center of the rotor of $41$ wind turbines.}
\end{figure}

In an accompanying domain $x\le-8D$, we employed Eq~(\ref{eq:lti}) to simulate an ensemble of eddies -- attached to the wall ({\em i.e.} Earth's surface) -- whose centers are randomly but uniformly distributed in space, and whose axes of rotation are all aligned in the $x$-direction.

In the absence of wind turbines or any other disturbances, Fig~\ref{fig:q} shows the transition of initially quiescent flow to a turbulent state. Indeed, the visualization is consistent with the findings of~\citet{Townsend80} that eddies that are attached to a wall can maintain an enhanced level of variance by extracting energy from the background flow and passing it to the perturbation field. The attached eddies, injected by the linear model, are stretched by the logarithmic mean profile in the domain, where the  mean strain rate tensor -- normalized by the angular velocity $2(u_*z_0)/(\kappa z)$ -- is
$$
\tilde{\mathcal S} = \left[
  \begin{array}{ccc}
    0&0&1\\
    0&0&0\\
    1&0&0\\
  \end{array}
  \right]
$$
Two eigenvectors of $\tilde{\mathcal S}$ corresponding to nonzero eigenvalues $\lambda_1=1$ and $\lambda_2=-1$ are $[1/\sqrt{2},\,0,\,1/\sqrt{2}]^T$ and $[1/\sqrt{2},\,0,\,-1/\sqrt{2}]^T$, respectively.

\begin{figure}
  \begin{tabular}{c}
    \includegraphics[height=5.5cm]{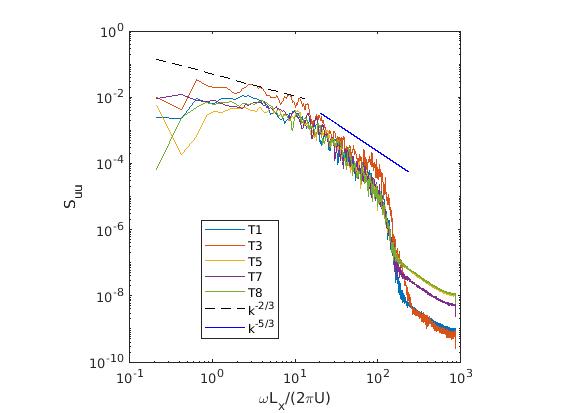}\\
    \includegraphics[height=5.5cm]{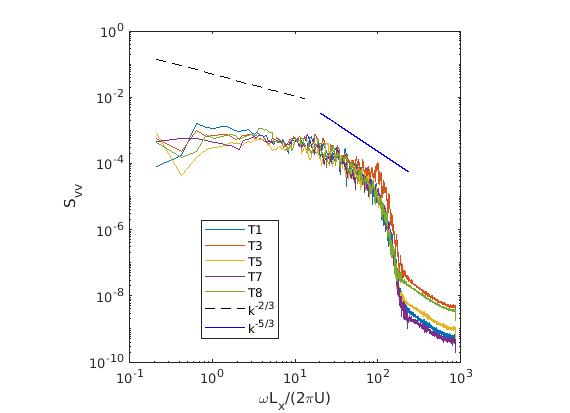}\\
    \includegraphics[height=5.5cm]{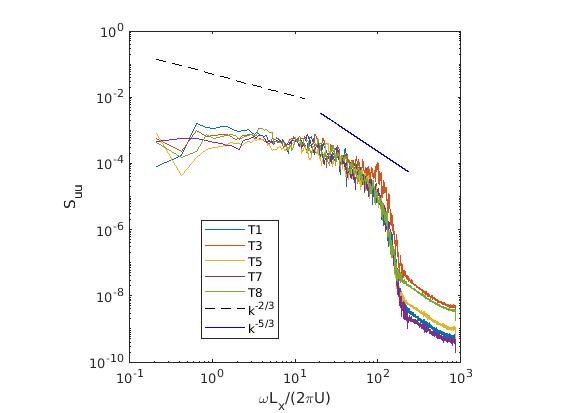}
  \end{tabular}
  \caption{\label{fig:uu} Auto spectral density functions of stream-wise, span-wise, and vertical velocity signals for five hub height locations, $z=90$, indicated with T1, T3, T5, T7, and T8 (see Fig~\ref{fig:prb}). The spectrum of the signals are compared with $k^{-2/3}$ and $k^{-5/3}$, where $k\propto\omega L_x/(2\pi U)$, $\omega$ is angular frequency, $L_x$ is horizontal length scale, and $U$ is the velocity scale. }
\end{figure}
\begin{figure}
  \begin{tabular}{c}
    \includegraphics[height=5.5cm]{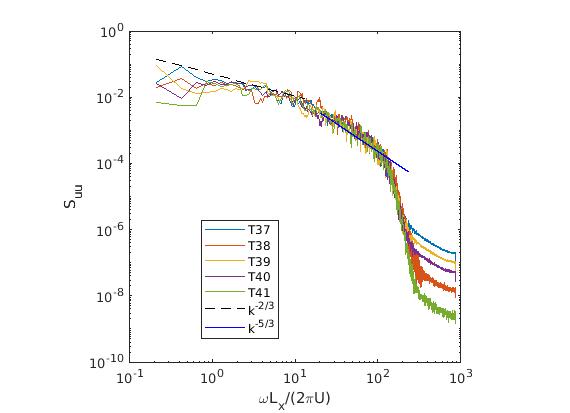}\\
    \includegraphics[height=5.5cm]{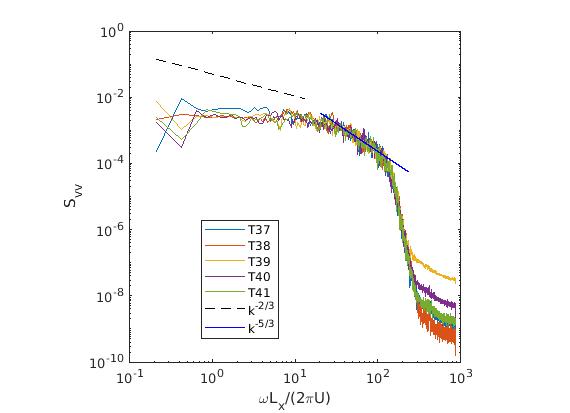}\\
    \includegraphics[height=5.5cm]{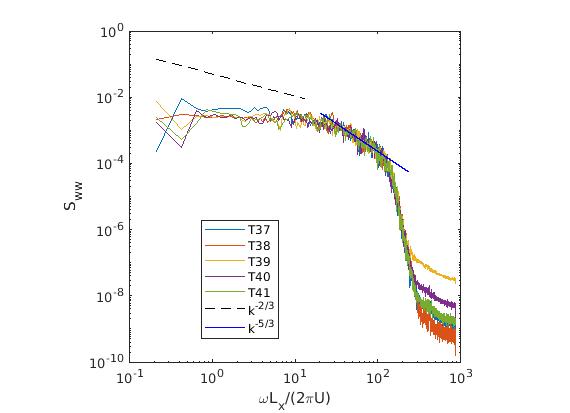}
  \end{tabular}
  \caption{\label{fig:lu} Similar to Fig~\ref{fig:uu}, auto spectral density functions of stream-wise, span-wise, and vertical velocity signals for five hub height locations, $z=90$, indicated with T37, T38, T39, T40, and T41 (see Fig~\ref{fig:prb}). }
\end{figure}

To critically analyze the transitional flow development, this study has sampled velocity signals at $41$ locations on the horizontal plane $z=90$~m, as indicated in Fig~\ref{fig:prb} (with T1-T41). The velocity signals were analyzed to understand the energy content of the eddying motion. In this simulation of the atmospheric boundary layer. For this analysis, consider the auto spectral density function (ASDF) defined by the Fourier transform of the autocorrelation
$$
\mathcal R_{uu}(t') = \frac1T\int_0^Tu(t)u(t+t')dt
$$
such as
$$
\mathcal S_{uu}(f) = \int_{-\infty}^{\infty} \mathcal R_{uu}(t')e^{-2\pi ift'}dt'.
$$
Since $\mathcal R_{uu}(t')$ is real and symmetric, setting $t'=0$ and applying inverse Fourier transform, we get
$$
\overline{u^2} = \int_\infty^\infty\mathcal S_{uu}(f)df,
$$
which indicates that $\mathcal S_{uu}(f)df$ represents the turbulence kinetic energy in a frequency band of $df$. 

Fig~\ref{fig:uu}-\ref{fig:lu} show auto spectral density functions ($S_{uu},\, S_{vv},\, S_{ww}$) for each of the three components of the velocity. We have compared $3$ velocity signals (T1, T3, T5) on the row indicated by `T1' in Fig~\ref{fig:uu}. and `T41', respectively; see in Fig~\ref{fig:prb}. A time-scale $L_x/U$ was considered to rescale the frequency, where the length $L_x = 56D$ accounts for $56$ rotor diameter and the velocity scale $U$ corresponds to the upstream wind at hub-height. From Townsend’s model of attached eddies, larger-than-inertial-scale coherent eddies can sense the effects of ground boundary as they are attached up to a displacement height from the surface~\cite{Townsend80}. These attached eddies are anisotropic and ``active'' in production of  shear  through the interaction between the subgrid scale momentum flux ($-\overline{u'w'}$) and the spanwise vorticity ($\partial U/\partial z$) of the mean flow. A comparison of the spectra with $k^{-2/3}$ indicates that attached eddies are in the range from Taylor microscale to approximately the height of the atmospheric boundary layer. A comparison of the spectra with $k^{-5/3}$ indicates a central tenet of the attached-eddy theory that the existence of a self-similar inertial subrange, where large eddies of size $D$ are detached from the ground surface ({\em i.e.} cannot sense the effect of the ground boundary).

In the atmosphere, and elsewhere, turbulent flows are characterized by apparently random and chaotic motions. The response to the nonlinear model from the linearized equation are not entirely unorganized in space and time. If the viscous terms were eliminated from the momentum principle, the resulting restricted Euler dynamics can lead to a singularity in finite time~\cite{Kang2020}. We know that the growth rate of the velocity gradient magnitude $||\bm\nabla\bm u||$ is driven by $-\rm{Tr}(\mathcal S\cdot\mathcal S\cdot\mathcal S) + (1/4)\bm\omega^T\cdot\mathcal S\cdot\bm\omega$. The growth of velocity gradient is opposed by the coherence of vortices as they align with the strain rate eigenframe. During such interactions, rapid rotation of vortices result in a turbulence stress on the surrounding strain field, and thus, the strain field must spend energy to oppose such stresses~\cite{Eyink2006,Hossen2021,Johnson2021}. The resulting autonomous dynamics of individual vortices display many traits that are observed in turbulent flows. Other than the relative alignment of vorticity vectors with the strain rate eigenframe, dynamical mechanisms that could advert the formation of singularities due to the growth of velocity gradient tensor are subgrid scale procecess and are not accessible in the LES methods. Recall the consideration of vortex stretching and alignment in Sec~\ref{sec:vst}.



\section{Spectrum of aerodynamic power fluctuations}
\label{sec:apf}
\begin{figure}[h]
  \centering
  \begin{tabular}{c}
    \includegraphics[height=4.5cm]{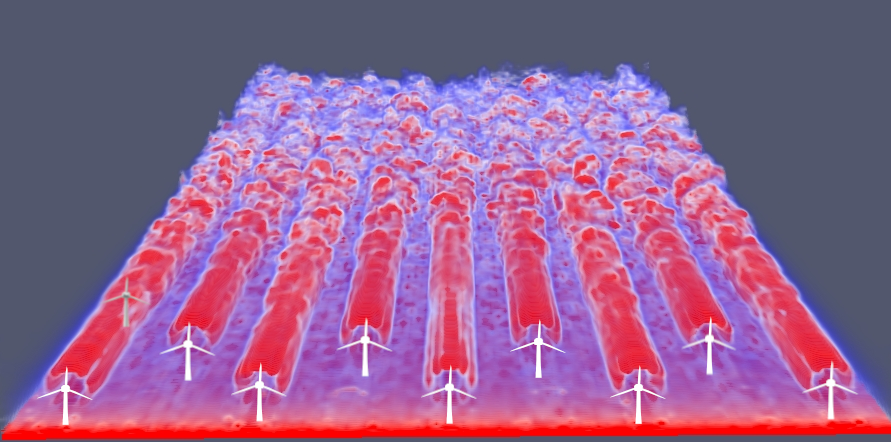}
  \end{tabular}
  \caption{\label{fig:vrt} An isosurface plot of the magnitude of vorticity ({\em i.e.} enstrophy) from a scale-adaptive LES of a wind farm.}
\end{figure}
Understanding the spectrum of wind power fluctuation in frequency $k$ as $k^{-5/3}$ is essential in managing the associated cost of wind power variability~\cite{Bandi2017,Bossuyt2017}. Furthermore, averaged fluctuations over geographically distributed wind farms exhibit a spectrum of $k^{-7/3}$.  This section connects of the wind power fluctuation spectrum with the stretching of vortices in wind turbine wakes. We have considered a large-eddy simulation of an idealized wind farm that consists of $41$ actuator disks with a rotor diameter of $126$~m, hub height of $90$~m, and thrust coefficient $C_t= 0.65\pm0.2$ (see~\citet{Xie2017}). Fig~\ref{fig:prb} shows the spatial locations of the center of $41$ turbines (T1-T41), where velocity signals were collected forming a covariance matrix. Fig~\ref{fig:vrt} shows the wind turbine wakes represented by the vorticity field.

A time signal of the thrust force  for $d$-th actuation disk is $F_d(t)\approx(1/2)\rho C_t\mathcal A\langle \bar u\rangle^2_d(t)$. The surrogate aerodynamic power signal is then $P_d(t)\approx (1/2)\rho C_t\mathcal A\langle \bar u\rangle^3_d(t)$. Here $\langle\bar u\rangle_d$ is the disk average of the LES-resolved velocity. 
To extract relevant features that capture invariant characteristics specific to each turbines, we have analyzed auto-correlation of the fluctuations in the aerodynamic power. For each signal, we performed the Reynolds decomposition $\langle\bm{\bar u}\rangle_d(t) = \tilde{\bm{\bar u}} + \bm u'(t)$. Substituting the decomposed velocity into the expression for instantaneous aerodynamic power ($\propto\langle\bar u\rangle_d^3$), we have the following cubic polynomial 
\begin{equation}
  \label{eq:pol}
P_d(t) = \frac12\rho C_t\mathcal A\left[\tilde{\bar u}^3 + 3\tilde{\bar u}^2u' + 3\tilde{\bar u}(u')^2 + (u')^3\right].
\end{equation}
In the above expression, we have adopted the streamwise component of the velocity $\tilde{\bar u} = \tilde{\bar u}_1$ for brevity. Neglecting fluctuating terms, we have the mean aerodynamic power
$$
\tilde P_d = \frac12\rho C_t\mathcal A\tilde{\bar u}^3.
$$
Thus, a linear correspondence between power and velocity fluctuations is
\begin{equation}
P'(t) \approx \frac32\rho C_t\mathcal A\tilde{\bar u}^2u'(t)
\end{equation}
if the higher order fluctuations are neglected. For typical marine atmospheric boundary layer flows over reasonably short time intervals ({\em e.g.} measurements at the Horns Rev wind farm~\cite{Barth2010}), it may be acceptable to neglect the higher order terms as first order approximation. 

However, intermittent gusting of wind may be large compared to typically expected order of magnitude of turbulence intensity.
To account for the higher order terms, the power fluctuation is calculated as
\begin{equation}
  \label{eq:pdp}
  P_d'(t) = P_d(t) - \tilde P_d,
\end{equation}
which accounts for all of the higher order terms in Eq~(\ref{eq:pol}).

\begin{figure}
  \centering
  \begin{tabular}{c}
    $(a)$\\
    \includegraphics[height=6cm]{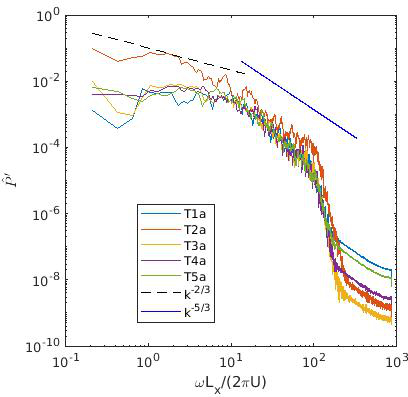}\\
    $(b)$ \\
    \includegraphics[height=6cm]{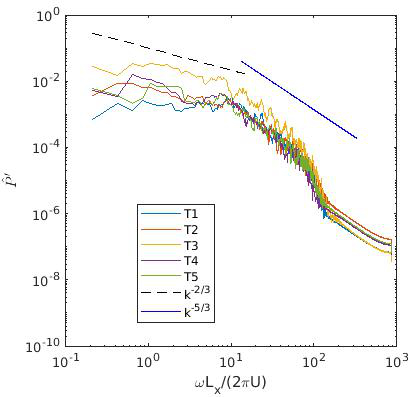}\\
    $(c)$ \\
    \includegraphics[height=6cm]{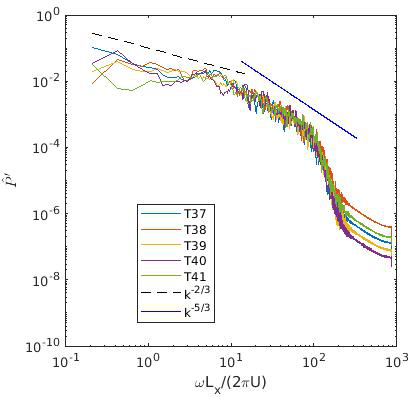}
  \end{tabular}
  \caption{\label{fig:p41} The spectrum of aerodynamic wind power fluctuations $P_d'(t)$; $(a)$ at five locations approximately $1.65D$ upstream to the first row of the turbines; $(b)$ at the first row the turbines; and $(c)$ at the last row of the turbines. }
\end{figure}
Plots of ASDF for power signals from $15$ wind turbines are shown in Fig~\ref{fig:p41}. It is worth mentioning that fluctuations in the aerodynamic power of individual wind turbines have been widely reported to follow the $-5/3$ spectrum of turbulence~\cite{Stevens2014,Bandi2017,Bossuyt2017}. The $-5/3$ spectrum of wind power fluctuation is associated with the small scale motion, order of $100$~m, that exists at a vertical layer where individual turbines operate~\cite{Bandi2017}.  Fig~\ref{fig:p41}$a$-$c$ show the spectral density of fluctuating power in the Fourier space, where the power autocorrelation $\hat P(f)$ is computed similar to the velocity autocorrelation $\mathcal S_{uu}(f)$ except for using $P'_d(t)$, Eq~(\ref{eq:pdp}). Here, the frequency $\omega/2\pi~\hbox{s}^{-1}$ has been scaled with the time scale $L_x/U$. At length scales $\mathcal O(1000$~m), turbulence is directly influenced by the largest length scales of atmospheric turbulence spanning hundreds of kilometers. For these simulated velocity signals, $L_x = 7056$ and $U=11$~m/s, which corresponds to a scaled frequency  $k=\omega L_x/(2\pi U)=1$. Fig~\ref{fig:p41} indicates that the large-scale fluctuations of the aerodynamic power, in a range of $700~\hbox{m}<L<7000~\hbox{m}$ (or $1<k<10$), scales like $k^{-2/3}$. At these length scales, turbulence seems to be associated with the clumpiness of vorticity and dissipation at a length that scales like the height of the atmospheric boundary layer. However, the small-scale fluctuations, in a range of $50~\hbox{m}<L<700~\hbox{m}$ (or $10<k<150$), exhibit a spectrum of $k^{-5/3}$. 

\subsection{Proper orthogonal decomposition}
\begin{figure}[t]
  \begin{tabular}{c}
    $(a)$\\
    \includegraphics[height=6cm]{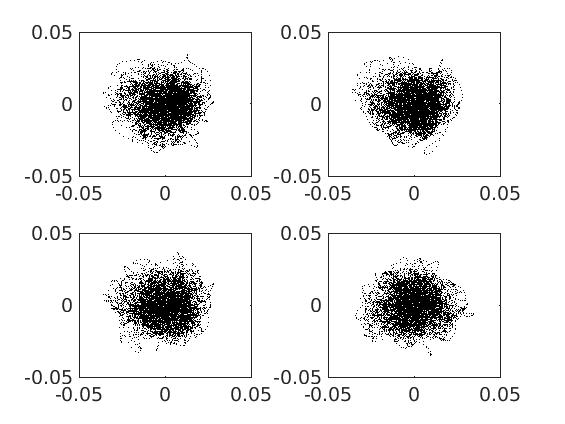}\\
    $(b)$\\
    \includegraphics[height=6cm]{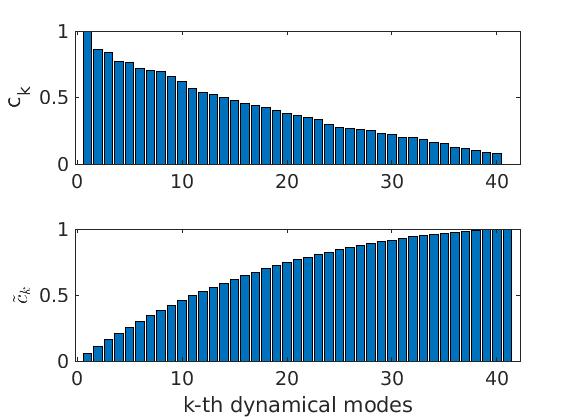}\\
  \end{tabular}
  \caption{\label{fig:pod} $(a)$ Scattered plots for pairs of turbines; from top left, counter clockwise, (T3, T7), (T3, T8), (T3, T12), and (T7,T8). $(b)$ Principal components of the POD modes for fluctuating aerodynamic powers of $41$ turbines.}
\end{figure}

The coherent structure of the atmospheric surface layer is characterized by ejections (of low-momentum eddies upward) and sweeps (of high-momentum eddies downward)~\cite{Garratt92}. In this section, we briefly study such a coherent structure using the proper orthogonal decomposition (POD) method. POD of complex turbulent flow fields extracts a low-dimensional coherent structure, which is a modal analysis methodology that is extensively used in data science ({\em albeit} under different acronyms). Briefly, a truncated decomposition of $u(x,t)$ with space-time separation is
\begin{equation}
  \label{eq:pod}
  u(\bm x, t) = \sum_{k=1}^rc_k(t)\phi_k(\bm x).
\end{equation}
A goal of the POD analysis is to identify a data-driven basis $\{\phi_k\}_{k=1}^r$ which represents the flow kinematics, while the time evolution of the coefficients $\{c_k(t)\}_{k=1}^r$ captures the low-dimensional coherent dynamics. The functions $\phi_k(\bm x)$ are  orthogonal, $\langle\phi_i,\phi_j\rangle = \delta_{ij}$, and they are called POD modes. The coefficients $c_k(t)$ are called principal directions of the time dynamics. The first coefficient vector $c_1(t)$ of the corresponding POD mode $\phi_1(\bm x)$ -- when sampled at discrete times -- represents the direction along which the temporal dynamics of turbulence fluctuates the most. The projection of $u(\bm x,t)$ along the direction of $c_2(t)$ to be uncorrelated with that along $c_1(t)$ is equivalent to constrain $c_1(t)$ to be orthogonal to $c_2(t)$. Thus, the principal components are linearly uncorrelated, $\langle c_i(t),c_j(t)\rangle = \sigma_i^2\delta_{ij}$, where $\sigma^2_i$ denotes the variability (or energy) of the signals decomposed by the POD method. 
\begin{figure}[h]
  \begin{tabular}{c}
    $(a)$\\
    \includegraphics[height=5.5cm]{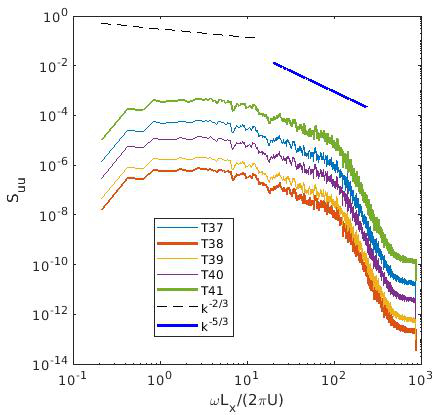}\\
    $(b)$\\
    \includegraphics[height=5.5cm]{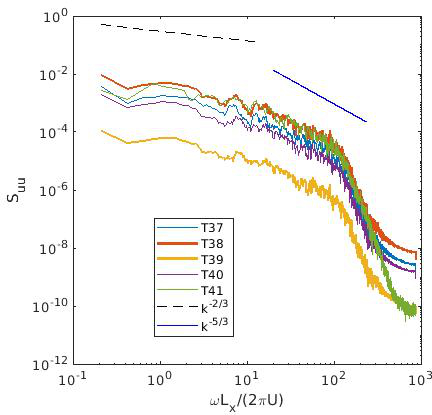}\\
    $(c)$\\
    \includegraphics[height=5.5cm]{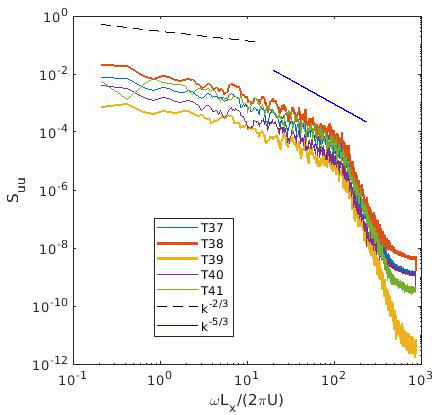}\\
  \end{tabular}
  \caption{\label{fig:rnk} ASDF of the aerodynamic power fluctuations  projected in the direction of the first POD. Corresponding velocity signals were taken at hub heights of the last row of turbines: T37, T38, T39, T40, T41. $(a)$~$r=1$, $(b)$~$r=3$, and $(c)$~$r=5$.}
\end{figure}

Consider a set of velocity signals $\{u(\bm x_d,t)\}$  corresponding to $n$ turbines, $d = 1,\ldots,n$. Here, we investigate the POD modes of the velocity fluctuation $u'(\bm x_d,t)$, thereby leading to a set of surrogate signals $\{c_k(t)\}$, $k=1,\dots,r\le m$. The decomposition is carried out by solving the following optimization problem:
$$
\operatorname*{\hbox{max}}_{\forall k, c_k(t)}\left[\frac{1}{n}\sum_{d=1}^n\left\langle c_k(t), u'(\bm x_d,t)\right\rangle^2\right]
\quad\hbox{s. t.}\quad\langle c_i,c_j\rangle = \delta_{ij},
$$
which is equivalent to the singular value decomposition (SVD) of the $n\times m$ snapshot of the velocity signals.

We have projected the fluctuating part of the aerodynamic power $P'_d(t)$, defined by Eq~(\ref{eq:pdp}), along the principal components of the POD modes. Fig~\ref{fig:pod}$a$ shows scattered plots of four pairs of the coefficients of POD modes: $[c_1(t),c_2(t)]$, $[c_1(t),c_3(t)]$, $[c_3(t),c_4(t)]$, $[c_3(t),c_5(t)]$. There is almost zero correlations among the POD coefficients, as expected. Fig~\ref{fig:pod}$b$ shows the principal components. We see that the first $10$ coefficients capture about $50$\% of the turbulence kinetic energy. 

Fig~\ref{fig:rnk} show the spectrum of wind power fluctuations for the five wind turbines on the last row of the wind farm. Here, we reconstruct the power fluctuations using Eq~(\ref{eq:pod}) for $r=1$, $r=3$, and $r=5$. Fig~\ref{fig:rnk}$a$ indicates that for $r=1$ the most variability of wind power fluctuations along the first principal direction is highly sensitive to self-attenuation of vortex stretching. As the number of POD modes $r$ increases, a trend in the convergence to $k^{-5/3}$ is observed. The spectrum of the wind power fluctuation does not vary in the presence of high non-linearity and vortex stretching  in wind turbine wakes.

\section{Conclusion and future direction}
\label{sec:cfd}
This article presents a scale-adaptive LES framework for wind energy applications. The accuracy of LES for studying atmospheric turbulence in wind farms depends on our ability to parameterize subgrid-scale turbulence fluxes and turbine-induced forces.  However, this work is dedicated mainly to the spectrum of wind power fluctuations in wind farms. The aggregated effect of a wind farm on the land-atmosphere exchange (of momentum fluxes) is adopted while considering the vortex stretching mechanism for the cascade of the energy flux. In the scale-adaptive LES framework, we have adopted one of the best-known mathematical results on the smoothness, regularity, and finite time blow-up of the solution of the Navier-Stokes equations. Thus, vortex stretching links directly the energy cascade and the smoothness of turbulent flow fields. More specifically, the scale-adaptive LES method employs the identities of~\citet{Betchov56} for developing a tuning-free dynamic subgrid model. This approach -- which seems to have been first illustrated by~\citet{Borue98} for homogeneous isotropic turbulence -- proves to help understand spectral characteristics of power fluctuations in wind farms.  Considering a stochastic dynamical model of inflow turbulence, we observe that large scale influence of the atmosphere may result in the $k^{-2/3}$ scaling, which is due to the frictional effects of the surface, while the $k^{-5/3}$ scaling persists in the inertial range without any effect from the surface. The POD analysis of the spectrum of wind power fluctuations shows that the local wake interactions influence the most variability of the power fluctuations of individual wind turbines. 

In closing this section, we note that for a complete understanding of the connection between vortex stretching and wind power fluctuations, one must also consider a more direct interaction of subgrid stresses with the filtered strain $\mathcal S$ and vorticity $\bm\omega$. One of the promising methods would be to apply the  wavelet decomposition as an explicit multi-scale filter forming the Leonard stresses~\cite{Alam2007,Alam2012,Alam2015,Park2016}. A potential of the wavelet method would achieve the scale-adaptivity hierarchically while capturing the most significant intermittent turbulence. This work is currently underway.

\section*{Acknowledge} Author acknowledges partial financial support from the Research Innitiative \& Services (RIS), Memorial University, and the National Science and Engineering Research Council (NSERC), Canada, in the form of a discovery grant. Author also acknowledges help from Mr. Jagdeep Singh, particularly in producing the plot shown in Fig~\ref{fig:vrt}.
\bibliographystyle{aipnum4-1}
\bibliography{refs}

\end{document}